\documentclass[5p,twocolumn,number,10pt,a4paper,sort&compress]{elsarticle}

\usepackage{ragged2e}
\makeatletter
\def\ps@pprintTitle{%
 \let\@oddhead\@empty
 \let\@evenhead\@empty
 \def\@oddfoot{{\RaggedLeft\tiny\copyright 2018. This manuscript version is made available under the CC-BY-NC-ND 4.0 license \url{http://creativecommons.org/licenses/by-nc-nd/4.0/}}}%
 \let\@evenfoot\@oddfoot}
\makeatother

\usepackage{subcaption}
\usepackage{amsmath,amssymb,amsthm}
\usepackage{graphicx}% Include figure files
\usepackage{dcolumn}% Align table columns on decimal point
\usepackage{bm}% bold math
\usepackage{hyperref}% add hypertext capabilities
\usepackage{siunitx}
\DeclareSIUnit\Molar{\textsc{M}}
\usepackage[capitalise]{cleveref}
\usepackage{multirow}

\usepackage[normalem]{ulem}
\usepackage{comment}
\usepackage[usenames,dvipsnames]{color}
\usepackage{soul}

\usepackage[mathlines,displaymath,switch]{lineno}% Enable numbering of text and display math
\newcommand*\patchAmsMathEnvironmentForLineno[1]{%
  \expandafter\let\csname old#1\expandafter\endcsname\csname #1\endcsname
  \expandafter\let\csname oldend#1\expandafter\endcsname\csname end#1\endcsname
  \renewenvironment{#1}%
     {\linenomath\csname old#1\endcsname}%
     {\csname oldend#1\endcsname\endlinenomath}}% 
\newcommand*\patchBothAmsMathEnvironmentsForLineno[1]{%
  \patchAmsMathEnvironmentForLineno{#1}%
  \patchAmsMathEnvironmentForLineno{#1*}}%
\AtBeginDocument{%
\patchBothAmsMathEnvironmentsForLineno{equation}%
\patchBothAmsMathEnvironmentsForLineno{align}%
\patchBothAmsMathEnvironmentsForLineno{flalign}%
\patchBothAmsMathEnvironmentsForLineno{alignat}%
\patchBothAmsMathEnvironmentsForLineno{gather}%
\patchBothAmsMathEnvironmentsForLineno{multline}%
}

\newcommand{\insilico}{\textit{in silico }}
\newcommand{\invitro}{\textit{in vitro }}
\newcommand{\invivo}{\textit{in vivo }}
\newcommand{\insilicons}{\textit{in silico}}

\newcommand{\changed}[1]{\hl{#1}}
\renewcommand\hl[1]{#1}
\renewcommand{\vec}[1]{\bm{#1}}
\newcommand{\intd}{\mathop{}\!\mathrm{d}}
\newcommand{\abs}[1]{\lvert{#1}\rvert}

\newcommand{\leish}{\textit{Leishmania\ }}
\newcommand{\lmex}{\textit{L.\ mexicana\ }}
\newcommand{\leishmex}{\textit{Leishmania mexicana\ }}
\newcommand{\leishns}{\textit{Leishmania}}
\newcommand{\lmexns}{\textit{L.\ mexicana}}
\newcommand{\leishmexns}{\textit{Leishmania mexicana}}

\soulregister\cite7
\soulregister\citet7
\soulregister\citep7
\soulregister\ref7
\soulregister\Cref7
\soulregister\cref7
\soulregister\subref7
\soulregister\si7
\soulregister\leish7
\soulregister\leishns7
\soulregister\lmexns7
\soulregister\lmex7
\soulregister\insilicons7
\soulregister\ 7

\begin{document}


\title{Boundary behaviours of \leishmexns: a hydrodynamic simulation study}
\author[wcmb]{Benjamin J. Walker\corref{cor1}}
\ead{benjamin.walker@maths.ox.ac.uk}

\author[dunn,nuff]{Richard J. Wheeler}
\ead{richard.wheeler@path.ox.ac.uk}

\author[wcmb,tokyo]{Kenta Ishimoto}
\ead{ishimoto@maths.ox.ac.uk}

\author[wcmb]{Eamonn A. Gaffney}
\ead{gaffney@maths.ox.ac.uk}

\cortext[cor1]{Corresponding author}

\address[wcmb]{Wolfson Centre for Mathematical Biology, Mathematical Institute, University of Oxford, Oxford, OX2 6GG, UK}
\address[dunn]{Sir William Dunn School of Pathology, University of Oxford, Oxford, OX1 3RE, UK}
\address[nuff]{Peter Medawar Building for Pathogen Research, Nuffield Department of Medicine, University of Oxford, Oxford, OX1 3SY, UK}
\address[tokyo]{Graduate School of Mathematical Sciences, The University of Tokyo, Tokyo, 153-8914, Japan}

\renewcommand*{\today}{15 November, 2018}
\date{15 November, 2018}% It is always \today, today,

\begin{abstract}

%!TEX root=../author_accepted.tex

It is well established that the parasites of the genus \leish exhibit complex
surface interactions with the sandfly vector midgut epithelium, but no prior
study has considered the details of their hydrodynamics. Here, the boundary
behaviours of motile \leishmex promastigotes are explored in a computational
study using the boundary element method, with a model flagellar beating
pattern that has been identified from digital videomicroscopy. In particular a
simple flagellar kinematics is observed and quantified using image processing
and mode identification techniques, suggesting a simple mechanical driver for
the \leish beat. Phase plane analysis and long-time simulation of a range of
\leish swimming scenarios demonstrate an absence of stable boundary motility
for an idealised model promastigote, with behaviours ranging from boundary
capture to deflection into the bulk both with and without surface forces
between the swimmer and the boundary. Indeed, the inclusion of a short-range
repulsive surface force results in the deflection of all surface-bound
promastigotes, suggesting that the documented surface detachment of infective
metacyclic promastigotes may be the result of their particular morphology and
simple hydrodynamics. Further, simulation elucidates a remarkable
morphology-dependent hydrodynamic mechanism of boundary approach, hypothesised
to be the cause of the well-established phenomenon of tip-first epithelial
attachment of \leish promastigotes to the sandfly vector midgut.

% \begin{description}
% \end{description}
\end{abstract}

\begin{keyword}
	Promastigote motility \sep Boundary element method \sep Flagellar beat \sep Low Reynolds number flow \sep \leishns-sandfly gut interaction
\end{keyword}

\maketitle

\section{Introduction}
\label{intro}
%!TEX root=../author_accepted.tex

\begin{figure}[t]
	\centering
	\begin{subfigure}[c]{0.15\textwidth}
		\centering
		\includegraphics[width = \textwidth]{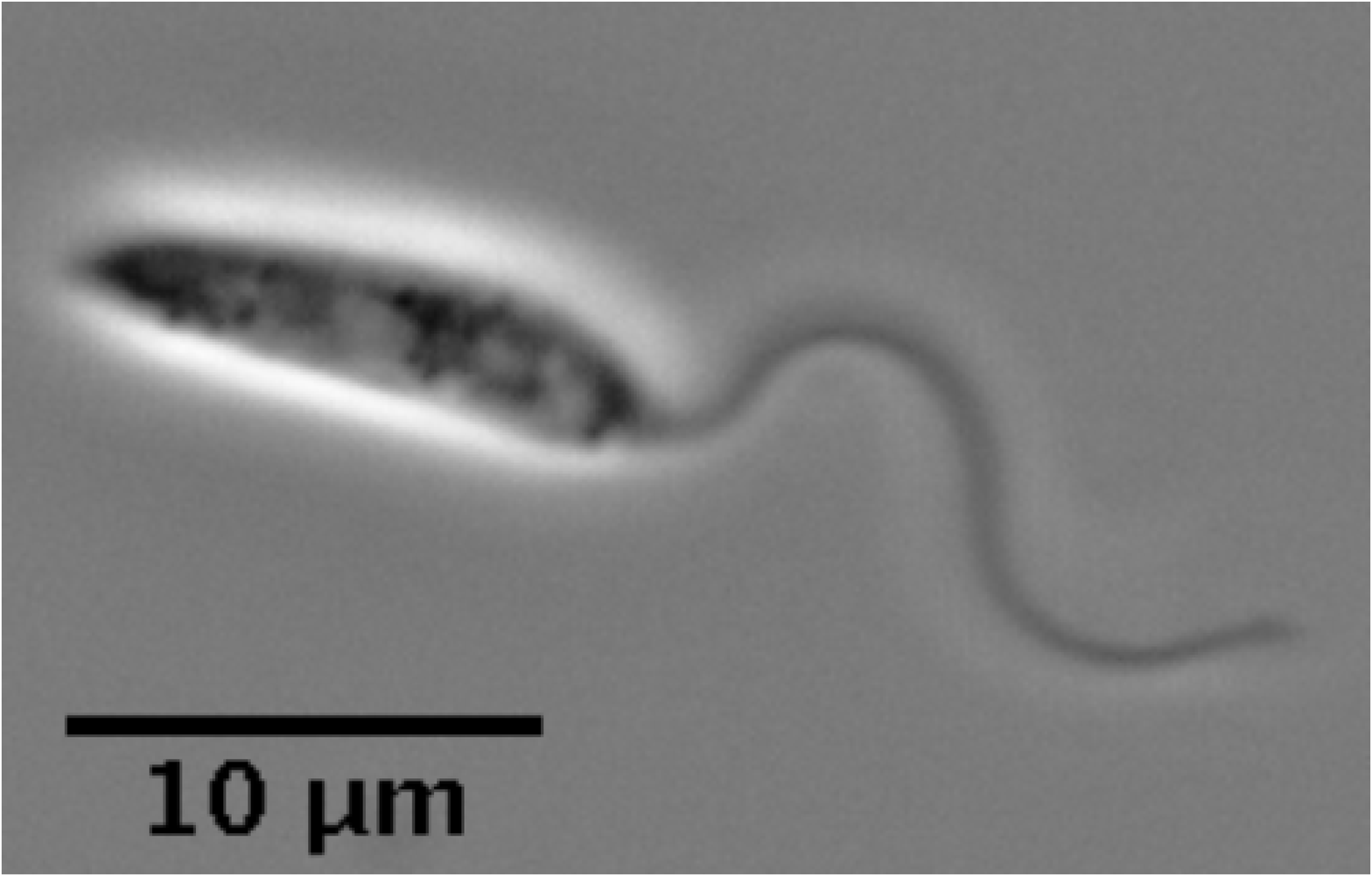}
		\caption{}
		\label{fig:leishmex_a}
	\end{subfigure}
	\begin{subfigure}[c]{0.15\textwidth}
		\centering
		\includegraphics[width = \textwidth]{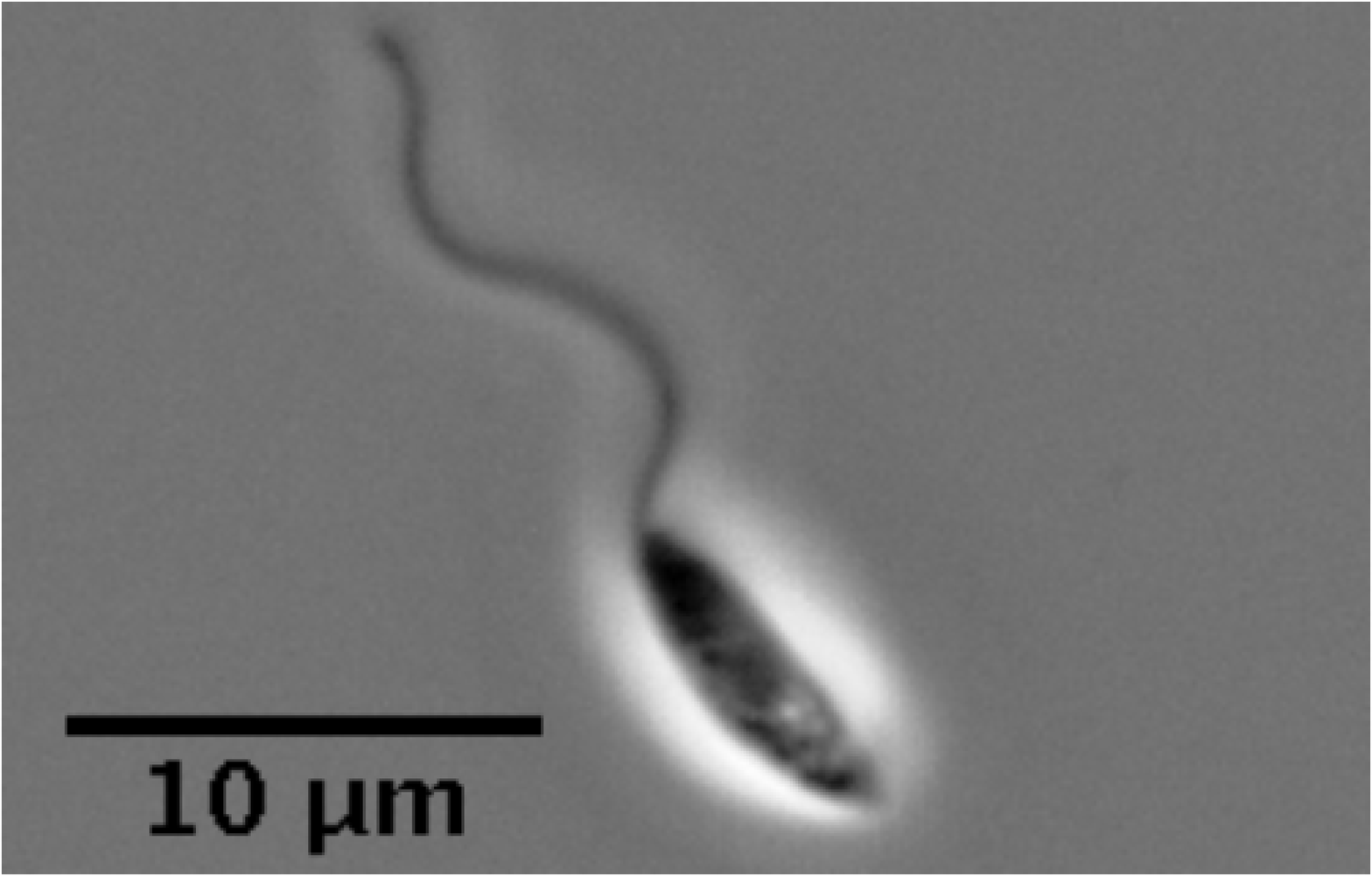}
		\caption{}
		\label{fig:leishmex_b}
	\end{subfigure}
	\begin{subfigure}[c]{0.15\textwidth}
		\centering
		\includegraphics[width = \textwidth]{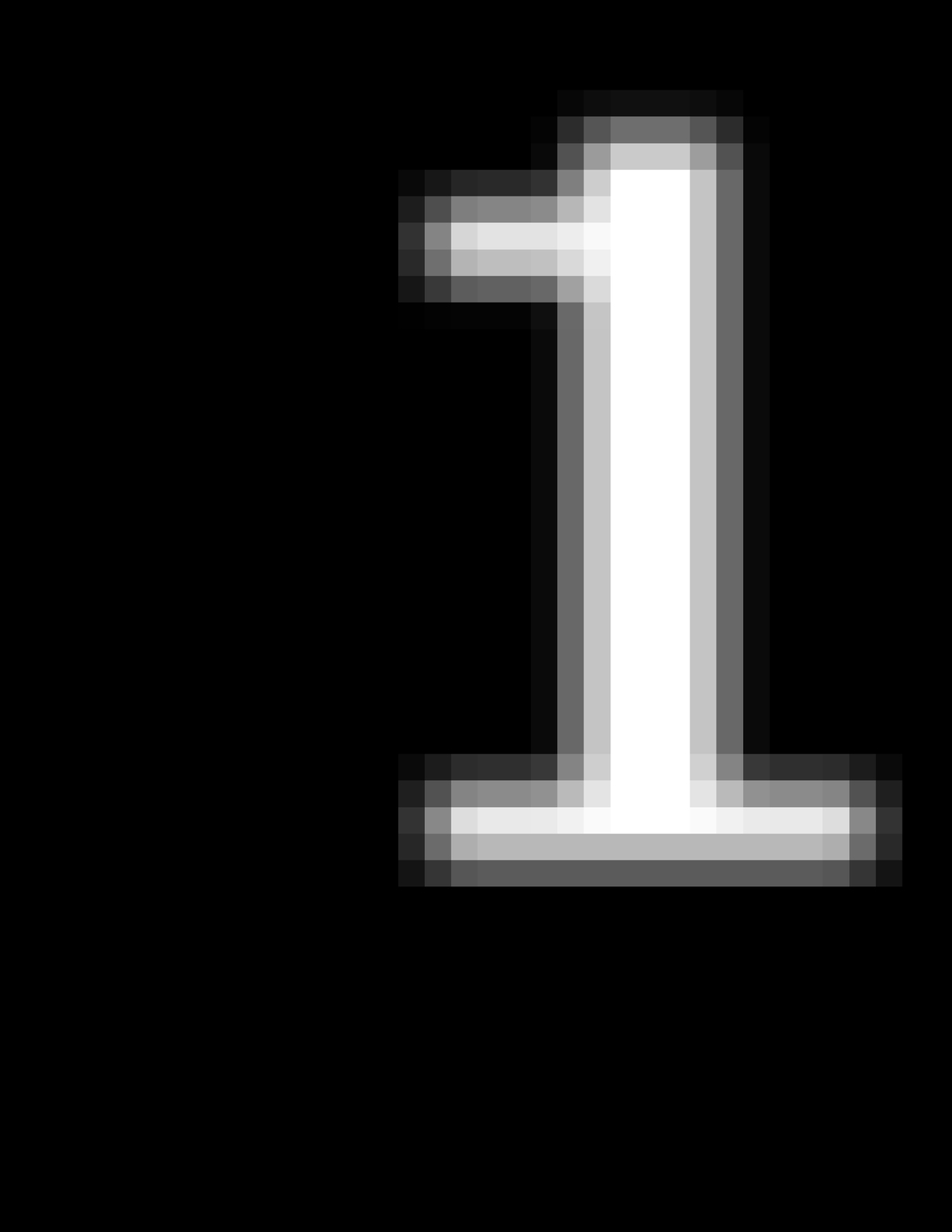}
		\caption{}
		\label{fig:leishmex_c}
	\end{subfigure}
	\caption{Sample frames from \lmex promastigote videomicroscopy,
	between two coverslips (\subref{fig:leishmex_a},\subref{fig:leishmex_b}),
	and in the bulk, as presented in montage form for
	(\subref{fig:leishmex_c}). We observe that the lengthscales of body and
	flagellum are approximately equal, with the cell body being approximately
	ellipsoidal in shape. Flagellar beating can be seen to be sinusoidal and
	planar in character, in both a confined environment and in the bulk.}
	\label{fig:leish_mex}
\end{figure}

The unicellular parasitic eukaryotes of the family Trypanosomatidae are the
cause of many major human diseases including African trypanosomiasis and New
World leishmaniasis \cite{Herwaldt1999}. Those of the genus
\leishns, transmitted to humans by the bite of a sandfly, affect
around 4 million individuals globally \cite{Herricks2013}. A prominent cause
of cutaneous leishmaniasis in the Americas, \lmex are a popular focus of
recent research owing to their complete development cycle being observable
\invitro \cite{Bates1994a}. In the highly motile promastigote stage of their
life cycle, a stage defined by morphology and as shown in
\cref{fig:leish_mex}, they utilise a single flagellum for locomotion,
protruding from their anterior cell body and predominantly beating with a
tip-to-base planar wave, the latter being common to all trypanosomatidae
\cite{Branche2006,gadelha2007a,
Goldstein1970,Holwill1974,Holwill1976a,Johnston1979b}. Their viability in the
sandfly vector midgut is thought to depend upon their ability to navigate
effectively \cite{Cuvillier2000a}, with it being widely accepted that their
survival in the low-Reynolds number environment of the sandfly midgut is
reliant upon attachment to the nearby epithelium
\cite{Dostalova2012,Bates2009}. In fact, the precise driving mechanism of the
tip-first boundary approach of \leish promastigotes remains unknown, and is
hypothesised by \citet{Bates2009} to simply be a naive consequence of their
flagellum-first swimming direction, but the effects of potential hydrodynamic
factors remain to be considered in detail. Contrastingly, in many
\leishns-sandfly pairings the mechanism of epithelial binding has been
well-explored, evidenced to be dependent upon the major
\leish surface glycoconjugate, \emph{lipophosphoglycan} (LPG)
\cite{Butcher1996,Pimenta1992,Pimenta1994,Soares2002,Soares2010}. Following
metacyclogenesis, and an accompanying change in LPG, the epithelial binding is
reversed, resulting in the detachment of the promastigote from the midgut
surface \cite{Pimenta1992}.

A direct consequence of locomotion via tip-to-base flagellar beating,
\leish spp.\ are hydrodynamically classified as \emph{pullers},
achieving propulsion by drawing fluid along the length of the flagellum before
then pushing out the fluid at the sides. This is in contrast to
\emph{pushers}, such as human spermatozoa and \textit{E. coli}, which perform
the reverse action and are consequently propelled in the opposite direction
\cite{Lauga2009}. Differences between the hydrodynamic properties of pushers
and pullers have been well documented for the case of \emph{squirmers},
swimmers of nearly constant shape with generated fluid flow at their boundary,
a model classically applied to \textit{Opalina} and other ciliated
microorganisms \cite{Chisholm2016,Ishimoto2013,Blake}. Further, within the
classes of pusher and puller fundamentally different behaviours are observed,
even for the simplest swimmers, as illustrated by the contrast between a
force-dipole puller, which deflects from boundaries \cite{Lauga2009}, and the
\changed{spherical} puller squirmer, which swims stably near boundaries
\cite{Ishimoto2013}. Hence refined models of cellular swimmers are required to
elucidate their boundary dynamics, as illustrated by the rich boundary
behaviours observed for flagellate pushers such as \textit{E. coli} and
mammalian spermatozoa, \changed{together with the biflagellate puller
\textit{Chlamydomonas}, in recent extensive work
\cite{Lauga2006,Lopez2014,Smith2009,Ishimoto2015,Fauci1995,Elgeti2011,Smith2011,Kantsler2013}}.
However, corresponding studies of monoflagellated pullers\changed{, either
observational or simulation-based,} are comparatively lacking and hence there
is extensive scope for the investigation of the boundary behaviours of a
flagellated puller such as \lmexns.

In particular, additional to their differing hydrodynamic classification,
\leish{}promastigotes are also morphologically distinct from the
better-studied pusher monoflagellates. Accumulation behaviours are
reported to be sensitive to variations in swimmer morphology
\cite{Ishimoto2017}, \changed{even for puller squirmers \cite{Ishimoto2013}}, while appeal to time reversal symmetry to infer puller
behaviour from pusher behaviour requires the same cellular morphology. Hence
\leish{}swimming behaviour cannot be inferred from previous studies of
swimmers, due to its distinct cell morphology, with the lengthscales of the
flagellum and cell body approximately equal (at approximately
$\mathop{\sim}10$\si{\micro\metre}, see \cref{fig:leish_mex}). In contrast,
for a typical human spermatozoon this ratio approaches one-tenth, with the
spermatozoon cell body being substantially shorter than the attached flagellum
\cite{Katz1986,Wheeler2011}.

The smaller cell body of such spermatozoa also enables the use of
approximate analytic techniques such as resistive force theory in studying
their motility \cite{Johnson1979}. This has been implemented classically for
the spermatozoa of the sea urchin \textit{Psammechinus} by \citet{Gray1955},
where hydrodynamic interactions between the two cell components are either
neglected or treated simplistically. Given the comparable scales of cell body
and flagellum in \leishns, such an approach is inappropriate
\cite{Johnson1979}, with methods treating the flagellum and cell body
comparably being more natural, and indeed, accurate. Thus, a full and
high-accuracy numerical study is necessitated to fully capture the
hydrodynamics and resulting behaviours of \leishns.

The functional relevance of a beating flagellum to the promastigote is the
control of spatial location. Hence we will examine the mechanics of
\leish upon approach to, and movement away from, a boundary. Thus
our aim is to consider how the cell may control its location in the sandfly
midgut, in its need to both approach and leave the gut epithelium at different
stages of its life cycle.

Hence, in this paper we will firstly detail digital capture for the flagellum
waveform of \leishmex in a typical growth medium. We then seek a
low-dimensional expansion of the observed kinematics via standard Fourier
analysis \cite{Werner2014,Ma2014}, and use a high-accuracy boundary
element computational framework to perform a number of \insilico experiments
\cite{pozrikidis2002}, our primary objective being to document the complex
long-term behaviour of a virtual promastigote in the presence of a planar
boundary. Further, we use beat-averaged phase planes to classify and quantify
behaviours in the near and far-field of a boundary, drawing simulation-based
behavioural comparisons with classical pushers such as the human spermatozoon.
Thus our final objective is to examine the hypothesis that hydrodynamic
interaction is sufficient for the tip-first boundary approach of \lmex promastigotes, and additionally that repulsive boundary interactions
drive the separation of promastigotes from a boundary.
\section{Methods}
\label{methods}
%!TEX root=../author_accepted.tex
\begin{figure}[b]
	\centering
	\includegraphics[width=0.4\textwidth]{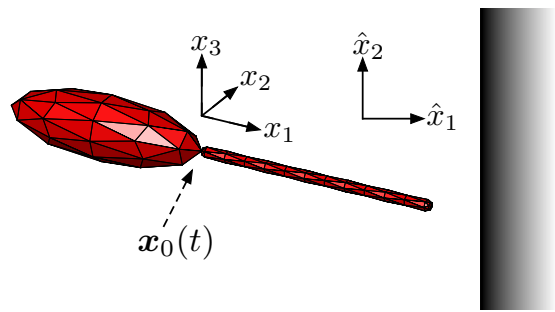}
	\caption{A 3-dimensional computational representation of the virtual
	promastigote with undeformed flagellum, where 80 triangular elements (red)
	have been used to mesh the cell body surface for illustration purposes
	(see \cref{app:mesh}). The location of the flagellar attachment at time
	$t$ is denoted $\vec{x}_0(t)$ in the laboratory frame, and is the origin
	of the cell-fixed reference frame whose axes, $x_1x_2x_3$, are depicted. A
	solid boundary lies in the laboratory frame plane
	$\hat{x}_1=\text{constant}$. (Colour online).}
	\label{fig:v_promast_ref_frame}
\end{figure}

\subsection{Videomicroscopy of \lmex}\label{sec:data_capture}
\leishmex high framerate videos were generated similarly to
previously described in \citet{Wheeler2017}. Promastigote \lmex (WHO
strain MNYC/BZ/62/M379) were grown in M199 supplemented with 10\% FCS and
50\si{\micro\Molar} HEPES$\cdot$HCl pH $7.4$, and maintained in exponential
growth between approximately $1\times10^6$ and $1\times10^7$ cells/\si{\ml}.
For high framerate videos, plain glass slides and coverslips were first
blocked by immersion in 1\% bovine serum albumen for 30\si{\s}, then washed
with distilled water. An approximately $2\times1$\si{\cm\squared} rectangle
was drawn on the blocked slide with a hydrophobic barrier pen, to which
1\si{\ul} \lmex culture in logarithmic growth was added, then a blocked
coverslip was added giving a ca.\ $\sim$5\si{\um} sample depth. 200
and 400 frame/\si{\s} videos between 4 and 9\si{\s} long were captured with an
Andor Neo v5.5 sCMOS camera using phase contrast illumination on a Zeiss Axio
Observer inverted microscope with a $63\times$ N.A. 1.3 Plan-Neofluar
objective (420881-9970-000) and a N.A. 0.55 long working distance condenser.
Sample frames can be seen in \cref{fig:leish_mex}. 

For visualising \lmex waveform in the bulk a 250\si{\um} thick adhesive
plastic square was applied to a glass slide to make a deep chamber,
in which 10\si{\ul} \lmex culture was placed, then a coverslip
added. Images were captured at a focal plane mid-way through the sample depth,
using dark field illumination and a long working distance $10\times$ N.A. 0.45
Plan-Apochromat objective (1063-139). Many cells lay with their cell body and
flagellum entirely in the focal plane, consistent with a planar flagellar beat
(see \cref{fig:leishmex_c}).

\begin{figure}[b]
	\centering
	\includegraphics[height = 0.2\textwidth]{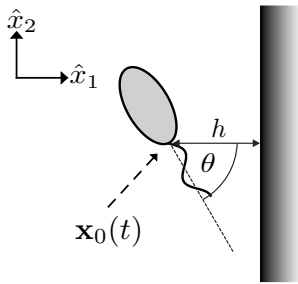}
 	\caption{Schematic showing the planar configuration of a virtual
 	promastigote, where boundary separation is denoted $h$, being measured
 	from the flagellum attachment point $\vec{x}_0$.\ The clockwise angular
 	displacement of the cell-fixed frame from the laboratory frame is denoted
 	$\theta$, as shown, such that $\theta=0$ corresponds to perpendicular
 	approach towards the boundary.}
 	\label{fig:def_of_h_theta}
\end{figure} 

\subsection{Determining flagellar kinematics}
Flagellar kinematics were extracted via automated analysis in the ImageJ macro
language, relative to a cell-fixed reference frame with coordinates
$\vec{x}=(x_1,x_2,x_3)$. This frame is defined as having $x_1$ directed along
the axis joining the body centroid to the visible base of the flagellum, with
the base being placed at the origin and having coordinates $\vec{x}_0$ in the
inertial laboratory frame (see \cref{fig:v_promast_ref_frame}). In a similar
analysis of mammalian spermatozoa \cite{Smith2009a,Ishimoto2014}, a tangential
attachment of the flagellum to the cell body was assumed due to the presence
of structural components, such as outer dense fibres, which provided
sufficient information to rotate the captured data into the cell-fixed frame.
Indeed, this would be appropriate for \leish at the true attachment site
inside the flagellar pocket, where microtubule structures bind the flagellum
to the cell body \cite{Lacomble2009,Lacomble2010}. However, our captured data
provides an exterior view at resolution such that the perceived flagellar
attachment appears free, and at a site that we will refer to as the
base, distinct from the true flagellar attachment zone in the flagellar
pocket. Therefore, relaxing the constraint of tangential attachment is
suitable here, and indeed this provides good agreement with observed
waveforms. Hence, the cell body rotation is instead used to determine the
cell-fixed frame orientation in the inertial frame, with the flagellar base
$\vec{x}_0$ being the centre of rotation here and throughout. Approximate
wavelengths and amplitudes were computed similarly to \citet{gadelha2007a},
where appropriate, along with an approximation to the cell body length. A
standard decomposition of the resulting waveform data into Fourier modes was
then performed \cite{Werner2014,Ma2014}, where an expansion of low dimension
was sought for use in numerical simulations.

\begin{figure*}[t]
	\centering
	\begin{subfigure}[c]{0.32\textwidth}
		\centering
		\scriptsize
		\begin{tabular}{cc}
			\hline
			Feature & Mean ($\pm$ S.D.)\\
			\hline
			Dominant frequency & $31.4\pm9.42$ \si{\hertz}\\
			% Freq.\ prominence & $45.8\pm24.9$\%\\
			Flag.\ wavelength & $4.21\pm1.00$\si{\um}\\
			Flag.\ amplitude & $1.77\pm0.877$\si{\um}\\
			Flag.\ length & $14.9\pm4.90$\si{\um}\\
			Body length & $9.81\pm3.29$\si{\um}\\
			Body width & 3.5\si{\um} \cite{Wheeler2011}\\
			Flagellum width & 0.3\si{\um} \cite{Wheeler2011}\\
			\hline
		\end{tabular}
		\caption{}
		\label{table:beat_cell_params}
	\end{subfigure}
	\begin{subfigure}[c]{0.32\textwidth}
		\centering
		\includegraphics[height = 0.6\textwidth]{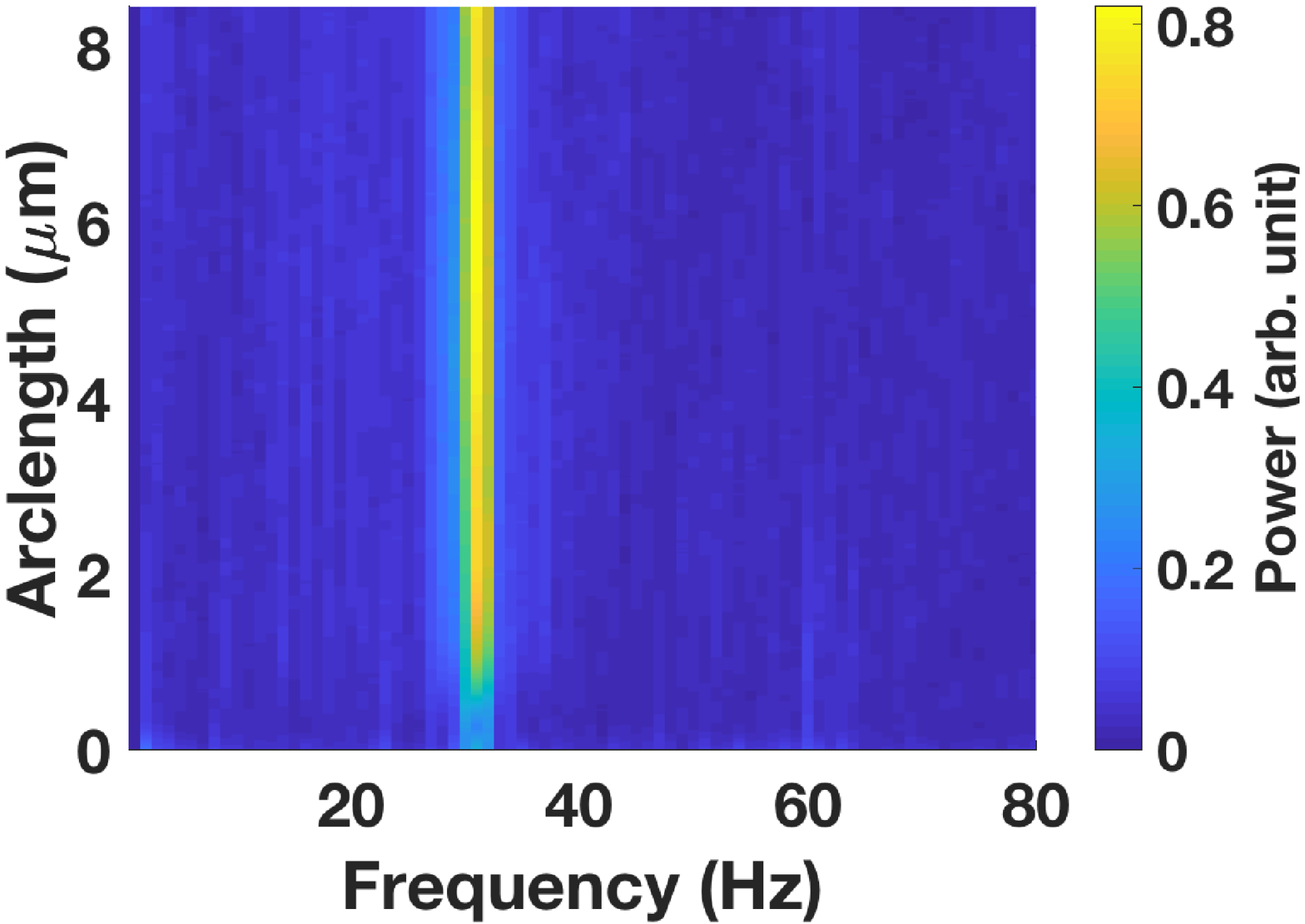}
		\caption{}
		\label{fig:fourier_spec}
	\end{subfigure}
	\begin{subfigure}[c]{0.32\textwidth}
		\centering
		\includegraphics[height = 0.6\textwidth,width=\textwidth]{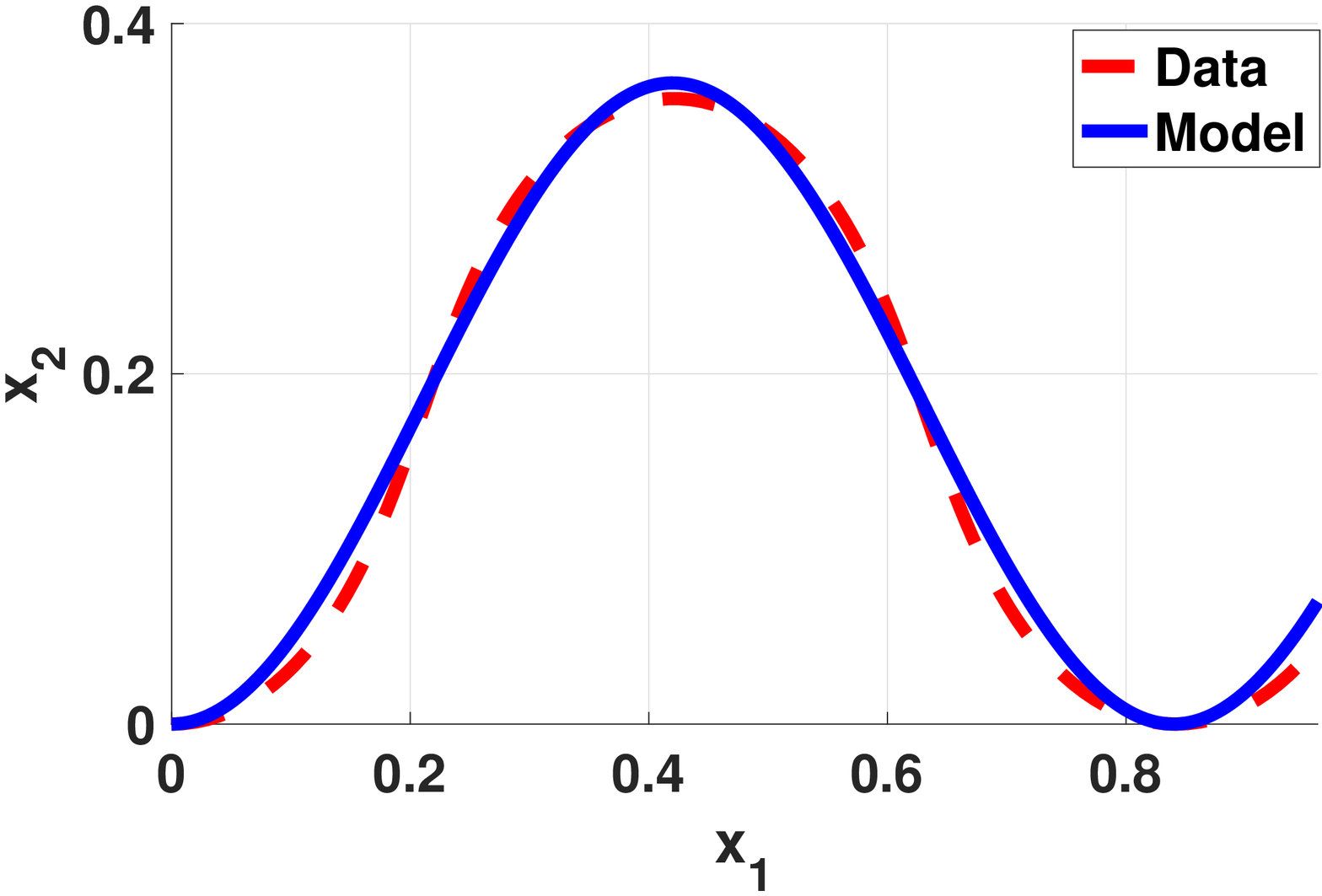}
		\caption{}
		\label{fig:model_beat_vs_data}
	\end{subfigure}
		\caption{(\subref{table:beat_cell_params}) Typical cell parameters as
		computed from an analysis of \lmex videomicroscopy of $N=126$ cells,
		along with body and flagellum widths from Wheeler et.\ al.
		\cite{Wheeler2011}. The flagellar beat amplitude is noted to be
		approximately half the typical cell body width. 
		% A high mean frequency
		% prominence is observed, where prominence is defined to be the
		% percentage difference between the median power of the two
		% most-powerful modes. Thus the prominence seen here supports the
		% hypothesis that \lmex flagellar dynamics are dominated by a single
		% mode. 
		(\subref{fig:fourier_spec}) The Fourier power spectrum for a
		sample cell. A strong dominant frequency band can be seen around
		28\si{\hertz}, present for the entire length of the flagellum. Power
		is shown here in arbitrary units, and arclength is measured from the
		flagellum base. (\subref{fig:model_beat_vs_data}) A flagellar waveform
		extracted from videomicroscopy (red, dashed) plotted against a model
		beat with a single sinusoid (blue, solid), showing excellent agreement
		both in space and in time (time not shown). (Colour online).}
\end{figure*}

\subsection{Governing equations} 
The small scale dynamics of \lmex promastigotes in a Newtonian fluid of
viscosity $\mu$ is governed by the incompressible Stokes equations (see
\cref{app:stokes}), with a Reynolds number on the order of $10^{-3}$, using
typical length and velocity scales given in \citet{Wheeler2011,Wheeler2017}.
For a given surface $S$, which typically will represent the promastigote
surface, we have the following non-dimensional integral representation for the
instantaneous flow velocity $\vec{u}$ relative to the inertial frame at a point
$\vec{x}^{\star}$ on the surface, with coordinates given in the body-fixed
frame and following \citet{pozrikidis2002},
\begin{equation}\label{eq:boundary_integral}
\begin{aligned}
	u_j(\vec{x}^{\star}) = & - \frac{1}{4\pi\mu}\int\limits_{S} G_{ij}(\vec{x},\vec{x}^{\star})f_i(\vec{x}) \intd{S}(\vec{x}) \\
	& + \frac{1}{4\pi}\int\limits_{S}^{PV} u_i(\vec{x})T_{ijk}(\vec{x},\vec{x}^{\star})n_k(\vec{x})\intd{S}(\vec{x})\,.
\end{aligned}
\end{equation}
Here, $G_{ij}$ and $T_{ijk}$ are velocity and stress Green's functions of
3-dimensional Stokes flow, $\vec{n}$ is the surface normal directed into the
fluid, $\vec{f}$ denotes the surface traction, and $\int^{PV}$ denotes a
principal value integral in the second summand. The velocity may be decomposed
into background, cell and disturbance velocities as in \citet{Ishimoto2014},
where the cell linear and angular velocities, $\vec{U}$ and $\vec{\Omega}$,
are a priori unknown in the inertial frame and must be solved along with the
boundary tractions, subject to additional force and torque-free constraints.

\subsection{The virtual promastigote}
\label{sec:v_promastigote}
In order to model the swimming behaviour of \leish we introduce a
neutrally-buoyant \textit{virtual promastigote}, which here will have an
idealised geometry that is similar to wild-type promastigotes (see
\cref{fig:v_promast_ref_frame,fig:leish_mex}). In particular, we construct our
idealised promastigote using an axisymmetric prolate ellipsoid to represent
the cell body, which differs slightly from observed \lmex promastigotes, as
the latter typically exhibit limited body curvature along their long axis (see
\cref{fig:leish_mex} for typical examples). With reference to the
non-dimensionalisation scales used in \cref{app:stokes}, we prescribe a
non-dimensional body length of $1.1$, with circular cross sections of diameter
0.35, consistent with a typical promastigote lengthscale of
10\si{\micro\metre} and corresponding to a non-dimensional flagellum length of
1.3. In turn, we model the latter by introducing a slender capped cylinder of
non-dimensional width 0.03 that attaches to the body at the origin of the
cell-fixed frame. The flagellum shape in the cell-fixed frame is described by
a general parameterisation
\begin{equation}
\begin{aligned}
	\vec{x} = \vec{H}(\xi,t)\,, \quad \xi\in[0,\xi^{\star}(t)]\,, \\
\end{aligned}
\end{equation} 
where the quantity $\xi^{\star}(t)$ is chosen such that the arclength of the
flagellum is conserved, and additionally we enforce that
$\vec{H}(0,t)=\vec{0}$, ensuring flagellar attachment occurs at the same
location on the body for all times $t$. Where it is appropriate to define a
beat plane in the cell-fixed frame, we will assume without loss of generality
that such a plane is spanned by unit vectors in the $x_1,x_2$-directions, so
that there is no beating in the $x_3$ coordinate direction (see
\cref{fig:v_promast_ref_frame}). Flagellum material velocities in the inertial
frame at a given time $t$ are approximated from positional information using a
central differences scheme optimised for double precision arithmetic
\cite{Nocedal1999}.

\subsection{Numerical scheme}
\label{sec:numerics}
Given the instantaneous velocity of the flagellum in the cell-fixed frame, we
proceed to solve the boundary integral equations of 3-dimensional Stokes flow
over the discretised virtual promastigote surface, closing the system with the
conditions of force and torque--free swimming, which are appropriate in the
inertialess limit of Stokes flow. Geometry, surface tractions and surface
velocities are interpolated using a mesh of $n$ nodes, yielding a linear
system of $3n+6$ equations in $3n+6$ unknowns, including the components of
swimming velocity $\vec{U}$ and angular velocity $\vec{\Omega}$ (for mesh
details see \cref{app:mesh}). We additionally enforce, without loss of
generality, that the normal boundary traction has a surface mean of zero,
eliminating the pressure non-uniqueness inherent in Stokes flow, and solve the
resulting system for the virtual promastigote velocities and surface
tractions. Throughout, we use the Blakelet for the integral kernel $G_{ij}$ in
\cref{eq:boundary_integral} \cite{Blake1971}, along with the accompanying form
of $T_{ijk}$, which ensures that the solutions satisfy a no-slip condition on
a specified planar boundary. Denoting coordinates in the laboratory frame by
$\vec{\hat{x}}=(\hat{x}_1,\hat{x}_2,\hat{x}_3)$, we typically specify this
stationary boundary as $\hat{x}_1 = 0$.

Having computed instantaneous promastigote velocities at a time $t$, we use
Heun's method with timestep $dt$ to update the position and orientation of the
cell in the inertial frame, as detailed in \citet{Smith2009}, with positional
evolution of the flagellum base point $\vec{x}_0$ following the
predictor-corrector scheme
\begin{equation}
 	\vec{x}_0(t+dt) = \vec{x}_0(t) + \frac{dt}{2}\left[\vec{U}(t) + \vec{U}(t+dt)\right]
\end{equation}
and cell orientation being dealt with similarly. An adaptive timestepping
scheme is employed in order to increase accuracy when approaching the
boundary, where velocities are expected to be highly sensitive to boundary
separation and body configuration.

Noting that surface interactions such as steric forces often occur
between cells and substrates, but also are highly variable between different
solutes and substrates \cite{Klein2003}, we proceed to additionally consider a
surface force. While an attractive surface force will simply tend to induce
binding once a cell is sufficiently close to a boundary wall, the impact of a
repulsive potential is ambiguous a priori, with the potential to reflect the
cell away from the boundary or to induce stable swimming for a cell that would
otherwise crash into the boundary. Hence we consider a repulsive boundary
force via a surface potential,  as introduced by \citet{Ishimoto2016a} and
motivated by Klein et.\ al's measurements \cite{Klein2003}. The resulting
force in non-dimensional form is given by
\begin{equation}\label{eq:potential}
	\vec{f}_{\text{wall}}(\vec{x}) = g\frac{e^{-d/l}}{1-e^{-d/l}}\vec{n}\,,
\end{equation}
where $d$ is the boundary separation, $\vec{n}$ is the outward-facing normal
of the boundary, and $l=0.02, \ g\propto\hat{\mu}/\hat{T}$ are the effective
range and strength of the force, chosen such that a strong short-range
repulsion is represented, with strength scaling with dimensionless viscosity
$\hat{\mu}$ and beat period $\hat{T}$.

Our implementation was verified in free-space against \citet{Ishimoto2014} and
by reproducing the Jeffery's orbits of ellipsoidal particles
\cite{Jeffery1922}, whilst the implementation of the Blakelet was compared
with the software library BEMLIB \cite{pozrikidis2002}.

\subsection{Construction of phase planes}\label{sec:phase_plane}
In an effort to gain a more complete picture of the virtual promastigote
dynamics without performing numerous costly individual simulations, we attempt
to simplify the dynamics via its restriction to a plane autonomous system, as
shown parameterised in \cref{fig:def_of_h_theta}. Specifically, we proceed by
equating the third coordinate vectors of the inertial and cell-fixed frames,
so that motion and beat plane are confined to the plane $\hat{x}_3=0$, without
loss of generality. We additionally average over a single beat period, as in
\citet{Ramia1993}, enabling a parameterisation by boundary separation and
orientation alone, which we denote by $h$ and $\theta$ respectively (see
\cref{fig:def_of_h_theta}). We define the separation to be the distance from
the flagellar attachment point to the wall, and the orientation to be the
clockwise angle between the body-fixed $x_1$-axis and the boundary normal. We
can then form the representation
\begin{equation}\label{eq:phase_plane_system}
	\begin{aligned}
 		\dot{h} &= F(h,\theta) \,, \\
 		\dot{\theta} &= G(h,\theta)\,, \\
	\end{aligned}
\end{equation}
where $F$ and $G$ represent the process of boundary element simulation and
subsequent phase averaging. Note that we may identify $-\dot{h} \equiv
\bar{U}_1$ and $-\dot{\theta} \equiv \bar{\Omega}_3$, the phase-averaged
linear velocity in the $\hat{x}_1$-direction and the rate of rotation about
the $\hat{x}_3$-axis respectively.
\section{Results}
\label{results}
%!TEX root=../author_accepted.tex

\subsection{\lmex exhibit simple flagellar kinematics}
Analysis of the temporal Fourier spectra of the \lmex flagellar beat for a
sample of $N=126$ cells between two cover slips revealed a single prominent
planar beat frequency in the range of 26-34\si{\hertz}, clearly observable
along the entire flagellum length (see \cref{fig:fourier_spec}). The lack of
other significant modes suggests a decomposition into a single sinusoid is
appropriate for representing the flagellar beat. Thus we opt to define the
non-dimensional beat parameters of amplitude, wavelength and frequency,
denoted $A$, $\lambda$ and $f$ respectively, and assume the functional form
\begin{equation}\label{eq:flag}
\begin{aligned}
	x_1(\xi,t) &= \xi \,,\\
	x_2(\xi,t) &= A \left[\sin{\left(\frac{2\pi}{\lambda}\xi + 2\pi ft\right)} - \sin{\left(2\pi ft\right)}\right]\,,\\
	x_3(\xi,t) &= 0
\end{aligned}
\end{equation}
in the cell-fixed frame. Under this assumption, amplitudes and wavelengths
were approximated, with the averaged results being shown in
\cref{table:beat_cell_params}. Here we set $A=0.18,\
\lambda=1.3$ and $f=2.8$ for use in simulations, recovering a typical
long-wavelength flagellar beat (see Supplementary Movie 1) and noting that the
results that follow are not sensitive to variations in these parameter
choices. Good agreement between the model flagellar beat and that extracted
from data is shown in \cref{fig:model_beat_vs_data}, demonstrating a
remarkably simple flagellar kinematics, not dissimilar to that of \textit{L.
major} and the classically-studied \textit{Crithidia oncopelti}
\cite{gadelha2007a,Holwill1974}. Additionally, it is noted that whilst this
planar beat pattern was seen in confined promastigotes (see
\cref{sec:data_capture}), such beating is also observed in the bulk and no
non-planar beating is exhibited (see \cref{fig:leishmex_c}). Furthermore, from
analysing observations of the human spermatozoon, it has  also been reported
that monoflagellate beating is unchanged near to a boundary to the resolution
that can be observed with typical microscopy \cite[Appendix A]{Ishimoto2014}.
Therefore we adopt our model beat pattern both in the far and near-field of
boundaries, and assume that there is not significant variation in this
waveform due to hydrodynamic boundary effects.

\begin{table}[b]
	\scriptsize
	\centering
	% \begin{tabular}{cccccccccc}
	\begin{tabular}{ccS[table-format=1.1]S[table-format=1.1]S[table-format=1.1]S[table-format=1.1]S[table-format=1.1]S[table-format=1.1]S[table-format=1.1]S[table-format=1.1]}
	&&\multicolumn{4}{c}{$\theta/\pi$} & \multicolumn{4}{c}{Body length}\\
	\rule{0pt}{1\normalbaselineskip}
	 & & \multicolumn{1}{||c}{0.2} & 0.3 & 0.4 & 0.5 & \multicolumn{1}{||c}{0.11} & 0.44 & 0.77 & 1.1 \\
	\cline{2-10}
	\rule{0pt}{1\normalbaselineskip}
	\multirow{4}{*}{$h$} & 0.6 & \multicolumn{1}{||c}{$\cdot$} & \multicolumn{1}{c}{$\cdot$} & 11.0 & 13.5 & \multicolumn{1}{||c}{2.8} & 3.9 & 6.9 & 11.0 \\
									& 0.8 & \multicolumn{1}{||c}{$\cdot$} & 8.7 & 7.2 & 8.3  & \multicolumn{1}{||c}{1.4} & 2.2 & 4.3 & 7.2 \\
									& 1.0 & \multicolumn{1}{||c}{$\cdot$} & 5.1 & 5.1 & 5.7 & \multicolumn{1}{||c}{0.9} & 1.5 & 3.0 & 5.1 \\
									& 1.2 & \multicolumn{1}{||c}{3.5} & 3.9 & 3.9 & 4.1 & \multicolumn{1}{||c}{0.6} & 1.1 & 2.3 & 3.9 \\
	\cline{2-10}
	\end{tabular}
	\caption{Percentage increase in hydrodynamic drag between the
	bulk and wall-facing sides of a virtual promastigote moving along the
	axis of the flagellum. Relative drag difference is shown for various
	configurations in phase space in the first section of the table for a
	virtual promastigote with typical morphology as described in
	\cref{sec:v_promastigote}, with the drag differential being
	greatest for low separations $h$. The dependence of this drag difference
	on swimmer body morphology is captured by the second part of the table,
	where we fix $\theta/\pi=0.4$ and vary the length of the swimmer body
	whilst preserving its aspect ratio. Increased body lengthscale is seen to
	significantly increase the relative drag difference, and thus a strong
	dependence of the resultant forcing of the swimmer on body morphology is
	highlighted. Boundary separation $h$ and body length are non-dimensional,
	and configurations that result in intersection with the boundary are
	omitted.}\label{tab:drag}
\end{table}

\begin{figure}[t]
	\centering
	\includegraphics[width=0.35\textwidth]{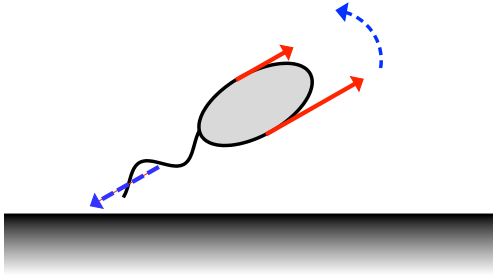}
	\caption{Physical mechanism of morphology-dependent drag-based rotation.
	The no-slip boundary induces increased drag on the nearside portion of the
	cell in comparison to the far side (red, solid), generating a net torque
	which causes cell reorientation, in addition to the typical promastigote
	movement towards the boundary (blue, dashed). (Colour online).}
	\label{fig:force_diagram}
\end{figure}

\begin{figure*}[t]
	\centering
	\begin{subfigure}[c]{0.45\textwidth}
		\centering
		\includegraphics[width=\textwidth,height=5cm]{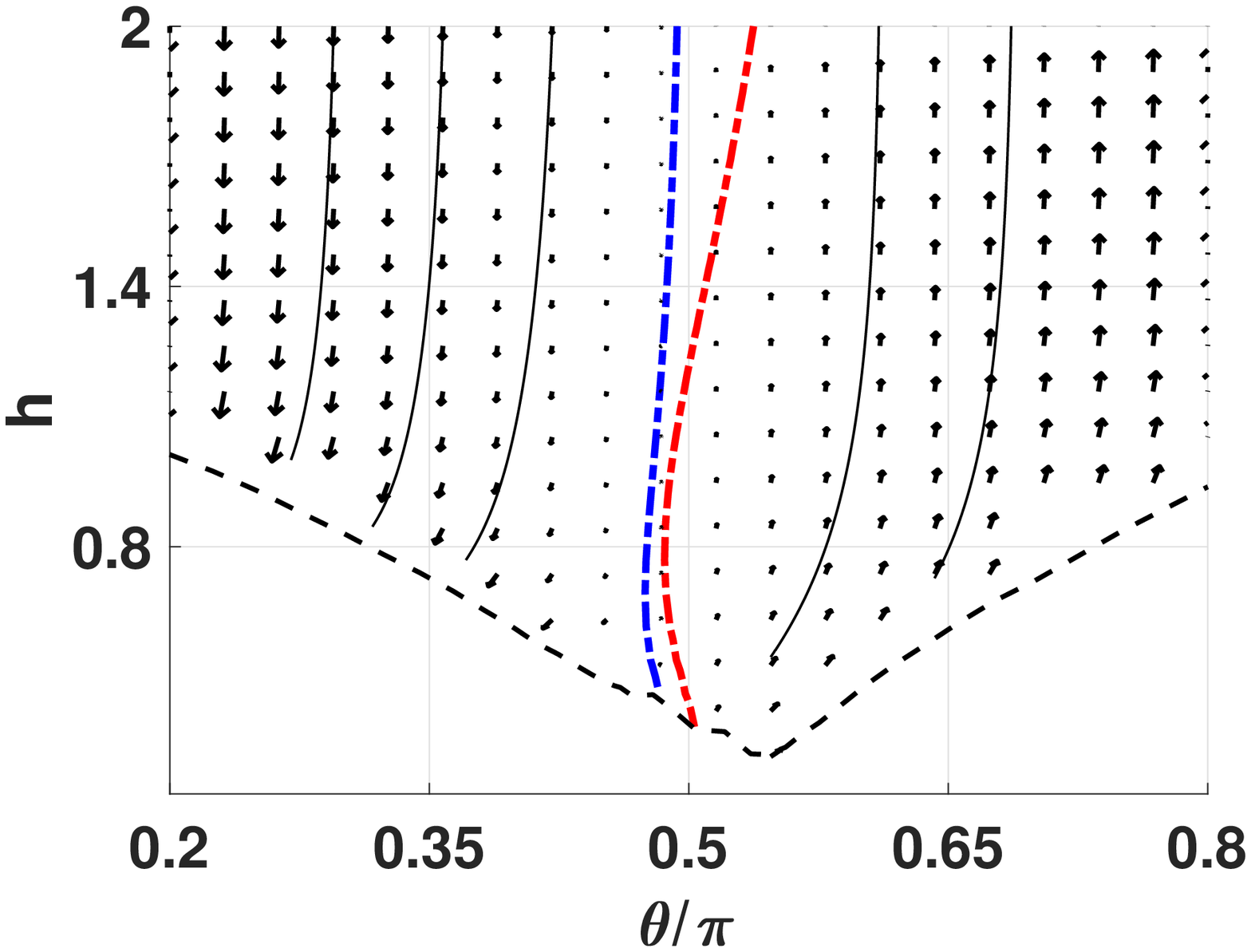}
		\caption{\label{fig:phase_plane_full}}
	\end{subfigure}
	\begin{subfigure}[c]{0.45\textwidth}
		\centering
		\includegraphics[width=\textwidth,height=5cm]{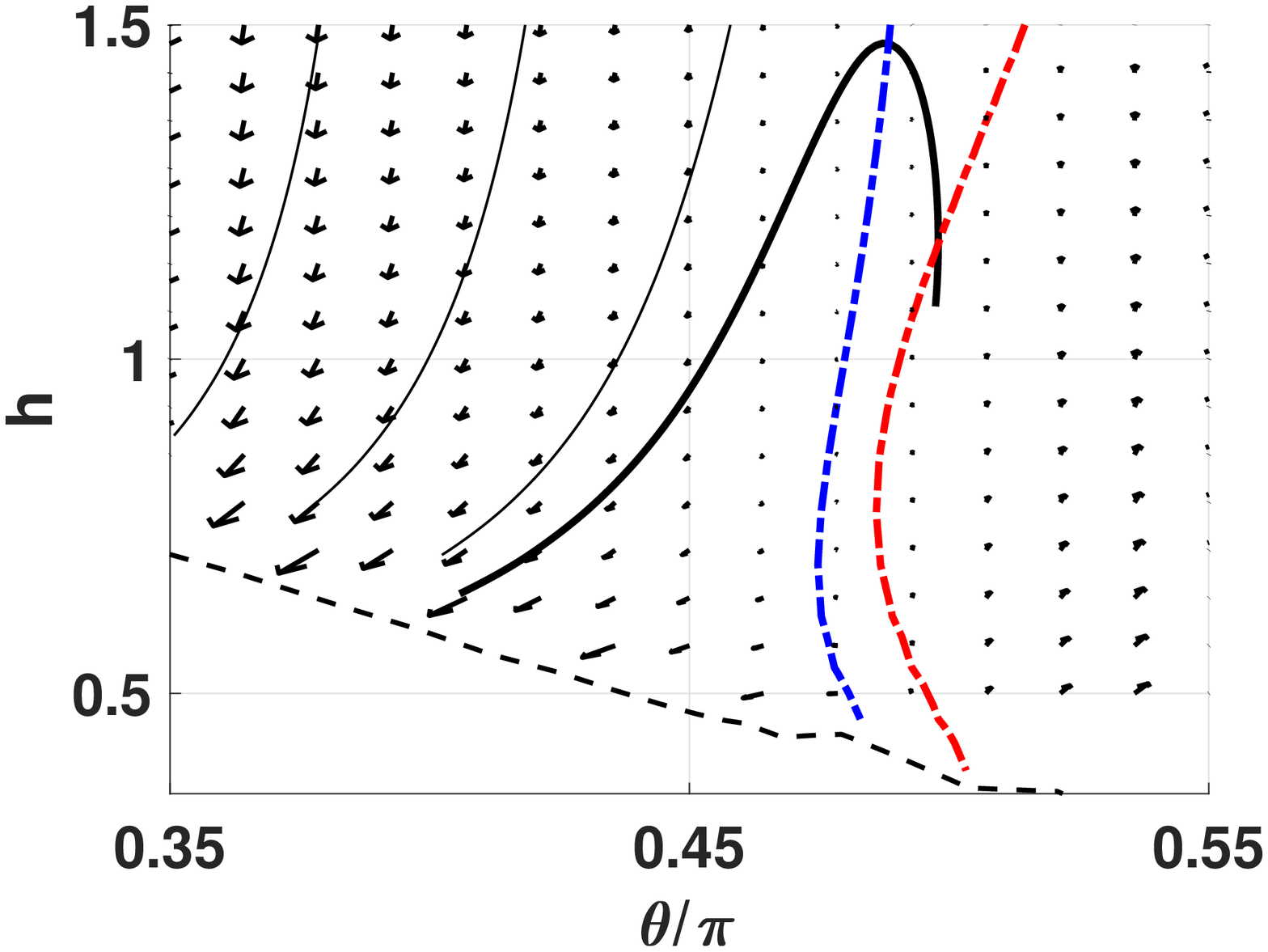}
		\caption{\label{fig:phase_plane_zoomed}}
	\end{subfigure}
	\caption{Beat-averaged phase planes approximating virtual promastigote
	motion near a boundary. Sample trajectories are shown as black lines, with
	the black dashed line separating off the region where configurations
	intersect with the boundary. Nullclines of the separation $h$ and
	orientation $\theta$ are shown as dashed lines (blue and red
	respectively), with the $h$ nullcline approaching $\theta=\pi/2$ as $h$
	becomes large. (\subref{fig:phase_plane_full}) We observe a full spectrum
	of dynamics, with boundary collision occurring in the left half of the
	phase plane, and reorientation away from the boundary in the right half.
	For cells approaching from the far-field ($h\rightarrow\infty$,
	$\abs{\theta}<\pi/2$), the trapping regions of the phase plane show that
	boundary collision occurs for most values of $\theta$, with deflection
	appearing possible only for $\theta\approx\pi/2$. The latter observation
	is confirmed by long-time simulation, where deflection may be observed if
	$\theta=\pi/2$ initially. The phase plane has no fixed points or periodic
	orbits, corresponding to a lack of stable boundary swimming.
	(\subref{fig:phase_plane_zoomed}) Higher resolution phase plane
	highlighting drag-based reorientation. Trajectories can be seen to curve
	rapidly in the direction of decreasing $\theta$ as they approach the
	boundary. A sample computed trajectory (heavy line) appears to exhibit a
	non-monotonic change in $h$. (Colour online).}
\end{figure*}

\subsection{Virtual promastigote beat plane aligns towards the perpendicular}\label{sec:perp_plane}
Long-time simulations of virtual promastigotes revealed a tendency to align
their beat plane normal to the boundary when $h$, the cell-boundary distance,
is sufficiently small. This may be explained by a simple torque balance
argument, especially due to the cell-body size, and exemplifies a general
observation that pullers tend to align perpendicular to boundaries
\cite{Lauga2009,Spagnolie2012}. In this instance, and with reference to
\cref{fig:force_diagram}, the no-slip boundary induces increased drag on the
near-side of the cell body in comparison to the far side. This difference
results in a torque, which when combined with the constraint of torque-free
swimming drives reorientation towards the perpendicular, significantly more so
than is present when the body size is reduced. The difference in hydrodynamic
drag across the virtual promastigote may be explicitly computed by prescribing
an instantaneous cell velocity and orientation in place of the force-free and
torque-free conditions, the results of which are shown in \Cref{tab:drag} for
various swimmer configurations and body lengthscales. These figures
demonstrate the existence of a notable drag difference between near and far
sides of the swimmer, and in particular the dependence of this difference on
body size, supporting the conclusion that increased cell body size is a factor
in the hydrodynamic promotion of promastigote reorientation. This behaviour,
and likewise the observations that follow, are observed to be robust to small
changes in cell aspect ratio and beat parameters (see \Cref{app:morphology}).

\subsection{Virtual promastigotes reorient to promote boundary collision via distal flagellar tip}
Upon a collision-bound approach to a planar boundary, in the absence of
additional repulsive surface forces, we observe the remarkable reorientation
of the virtual promastigote such that the distal tip of the flagellum is
promoted as the point of contact. This is visible in numerous individual
simulations, the behaviour also being captured by the phase plane
(\cref{fig:phase_plane_zoomed}), where we see a rapid change in angle $\theta$
when in close proximity to the boundary. This may again be partially explained
by the general tendency of pullers to align perpendicular to a wall,
but the large magnitude of the effect suggests that it is also resultant of
the same drag-based mechanism as the beat plane reorientation above. Indeed,
this is confirmed by repeating simulations with greatly-reduced body size,
where the effects are seen to be substantially decreased, consistent with the
quantitative observations shown in \Cref{tab:drag}. Thus both beat plane
orientation and the specifics of boundary approach are dependent upon virtual
promastigote morphology, and together result in the flagellar tip being the
primary point of surface contact.

\subsection{Swimming is unstable in the absence of a surface force}
As described in \cref{sec:phase_plane}, we compute a phase plane in order to
deduce long-time behaviour (see \cref{fig:phase_plane_full}). Partitioning
phase plane trajectories by the $h$ nullcline (which approaches $\theta=\pi/2$
in the far-field, depicted blue, dashed), we observe that almost all virtual
promastigotes approaching the boundary from the far-field will eventually
collide with the boundary (see Supplementary Movie 2). Similarly, almost all
trajectories that initially face away from the boundary exhibit a monotonic{}
increase in boundary separation and relative angle, with those that are
sufficiently close to the boundary undergoing deflection, where $\theta$
initially increases and then approaches a constant value as promastigote-wall
separation grows.

However, the phase plane shows a region where non-monotonic change in $\theta$
appears plausible (see \cref{fig:phase_plane_zoomed}), suggesting that those
promastigotes beginning on trajectories in the region between the $h$ and
$\theta$-nullclines will initially move away from the boundary, but will
subsequently undergo reorientation towards the surface and eventually collide
with it. Contrastingly, such behaviours are not observed in full simulations,
however this is not a significant conflict given the magnitude of the error
introduced by the phase plane averaging process. This error is typically of
the order $10^{-6}$, comparable to the $\dot{\theta}$ values seen in this
limited region of the phase plane near the nullclines, therefore the
approximate behaviour need not accurately reflect the full dynamics local to
this region. Thus, overall we see a dichotomy of behaviours exhibited by the
virtual promastigote, those of collision with or deflection away from the
boundary, with no mechanism for stable boundary swimming. This is confirmed by
long-time simulation of virtual promastigotes with $\theta\approx\pi/2$
initially, which orient away from the boundary rather than swimming stably
(see Supplementary Movie 3).

\begin{figure}[b!]
	\centering
	\includegraphics[width=0.45\textwidth,height=5cm]{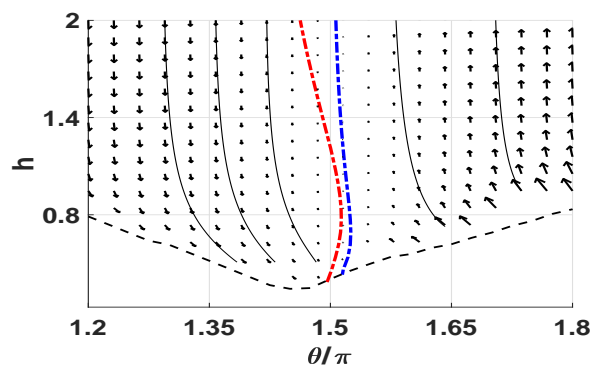}
	\caption{Beat-averaged phase plane approximating the motion near a
	boundary for a virtual pusher with the same morphology as the virtual
	promastigote of \cref{fig:v_promast_ref_frame}. Sample trajectories are
	shown as black lines, with the black dashed line separating off the region
	where configurations intersect with the boundary. Nullclines of the
	separation $h$ and orientation $\theta$ are shown as dashed lines (blue
	and red respectively), with the $h$ nullcline approaching $\theta=3\pi/2$
	as $h$ becomes large. Having reversed the propagation direction of the
	flagellar beat, we recognise the time-reversed behaviour of the virtual
	promastigote, subject to a shift in $\theta$ (cf.\
	\cref{fig:phase_plane_full}). Noting the sign of $\dot{\theta}$ near the
	$\theta$-nullcline, we see the stable attractor of $\theta\approx3\pi/2$.
	(Colour online).}
	\label{fig:phase_plane_pusher}
\end{figure}

In order to compare these behaviours against an appropriate pusher, we reverse
the direction of beat propagation in the flagellum and simulate the resulting
motion. The behaviour of this pusher is captured by a phase plane (see
\cref{fig:phase_plane_pusher}), where we observe that $\theta\approx3\pi/2$ is
a stable attractor of the system, corresponding to swimming parallel to the
boundary. In long-time simulations of the full system we in fact observe
stable boundary swimming at a constant separation $h$, which differs slightly
from the beat-averaged system due to the previously-described beat cycle
averaging errors near the system nullclines. In explanation of this stable
swimming with reference to the virtual promastigote, we firstly note that a
change in beat propagation direction is in fact equivalent to a reversal of
time in the flagellar kinematics of \cref{eq:flag}, subject to a phase shift.
As time appears only as a parameter in the governing equations, the solution
to the time-reversed problem is simply the reversal of the original problem.
Therefore we recover the reversed behaviour of the virtual promastigote in the
behaviours of this pusher, and hence observe stable parallel swimming in the
place of unstable motion. Additionally, we examine the effects of significant
decrease in cell body size by further modifying the virtual promastigote,
reducing the body volume by two orders of magnitude, and repeating the above
analysis. From this we note that the stable swimming of the altered pusher may
still be observed, albeit at a different boundary separation $h$, whilst the
puller behaviour remains unstable. Thus, our results suggest that hydrodynamic
classification is a more significant factor than morphology in determining the
stability of boundary swimming.

\subsection{Repulsive surface forces do not give rise to stable
boundary swimming} 
\begin{figure}[t]
	\centering
	\includegraphics[width = 0.44\textwidth]{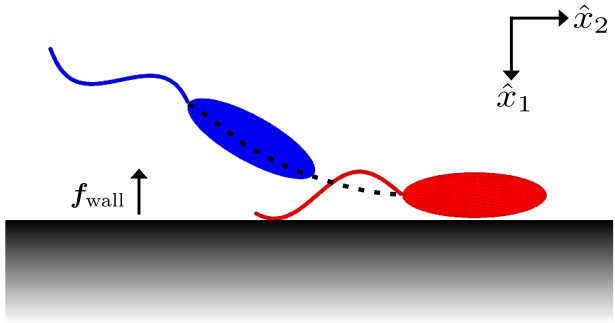}
	\caption{The simulated motion of a virtual promastigote in
	proximity to a boundary, accounting for a repulsive surface potential. The
	computational mesh is shown for the initial configuration (red), parallel
	to the boundary, and the configuration after ca.\ 1.4\si{\s} (blue). The
	approximate time-averaged path of the attachment point is also shown
	(black, dashed). We observe that the promastigote orients away from the
	boundary, and thereafter moves off into the bulk. See Supplementary Movie
	4 for full motion. (Colour online).}
	\label{fig:longtime_traj}
\end{figure}

The repulsive surface potential of \cref{sec:numerics} is
introduced to the system. Due to the short range over which the resulting
repulsive force is non-negligible, the previous phase plane analysis that does
not account for the surface force holds throughout most of the phase space.
However, due to the varying distance of the flagellum from the boundary
throughout a single beat period, phase averaging is not appropriate to
determine the near-boundary dynamics. Hence numerous long-time simulations
were performed (see \cref{fig:longtime_traj} for an example) to ascertain the
motion of the virtual promastigote when in close proximity to the boundary,
following which the phase plane of \cref{fig:phase_plane_full} is used to
examine further behaviour.

One might envisage the existence of a periodic motion, with the torque induced
by the surface potential orienting the cell into a configuration where it will
again collide with the boundary, and such motion repeats ad infinitum. This
phenomenon could not be observed for all conducted \insilico experiments with
typical parameter values and cell configurations, which we reason is due to
the strong repulsive boundary character.

We observe that the only exhibited behaviour is deflection away from the
boundary, with the surface repulsive force causing reorientation of the
promastigote such that it falls into the deflective regime, which is observed
to be unchanged for reasonable variation in force strength due to the
short-range nature of the surface potential. Thus the inclusion of a repulsive
surface potential of physically-reasonable character to model boundary surface
forces does not give rise to stable boundary swimming of virtual
promastigotes, and instead promotes their eventual deflection.

Reversing the direction of beat propagation we examine the near-boundary
behaviour of the morphologically-equivalent pusher, which due to the repulsive
surface force may not be inferred by time reversal of the puller behaviour. In
contrast to the deflection observed for the puller, in this case we in fact
see that the swimmer aligns approximately parallel to the boundary,
approaching a stable separation from the wall after an initial period of
transition, qualitatively similar to the non-deflective behaviour of
small-bodied pushers and as exemplified in Supplementary Movie 5. Thus our
results further evidence the significance of hydrodynamic classification in
determining boundary behaviours, and suggest in this case the subdominance of
body morphology in effecting stable boundary swimming.
\section{Discussion and conclusions}
\label{disc_conc}
%!TEX root=../author_accepted.tex

In this work we have identified and described a novel model flagellar waveform
of \lmex promastigotes, observing a simple planar dynamics in $N=126$ cells
that lends itself to simple parameterisation and subsequent simulation. Using
this model beat pattern we have observed and explored the boundary behaviours
of a virtual axisymmetric promastigote, an idealised hydrodynamic puller with
a significant body lengthscale.

% First objective: determine whether Leish boundary accumulate in the absence of repulsive surface forces?

We have seen that swimming near a boundary in the absence of surface forces is
unstable for virtual promastigotes, with trajectories resulting either in
immediate deflection away from the boundary or eventual collision with the
surface. This behaviour may not be deduced from previous observations
of puller microswimmers, owing to the swimmer's distinct cell morphology and
the reported sensitivity of boundary behaviours to swimmer geometry
\cite{Ishimoto2017}. Here, promastigotes that initially swim away from a
boundary may be captured if sufficiently close, undergoing reorientation which
results in their collision with the surface. Drawing comparison with \invivo
promastigotes, this may provide hydrodynamic explanation for the epithelial
attachment observed in the sandfly midgut that precedes promastigote
metacyclogenesis, established to be necessary for the survival of \leish
promastigotes \cite{Bates2009,Dostalova2012}.

% Comparison to pushers/smaller cell bodies.
Further, we have noted that time-reversibility of the governing equations
allows us to immediately compare the dynamics of pushers and pullers near
boundaries in scenarios without repulsive surface forces. Whilst our results
have shown that no stable boundary swimming occurs in the case of our
\changed{virtual promastigote puller}, we observed the stable motion of
morphologically-equivalent pushers, with the latter swimming parallel to the
boundary. This is in agreement with the known behaviour of the smaller-bodied
human spermatozoa, a pusher that can maintain stable parallel swimming next to
a planar boundary \cite{Fauci1995,Smith2009,Smith2009b,Elgeti2011}. A
significant reduction in cell body size is not observed to drastically alter
the behaviour of the virtual pusher, with only the stable swimming height
being affected. Thus our results show that the stable boundary swimming of
monoflagellates in quiescent fluid is highly dependent on the method of
locomotion, and less so to specific changes in morphological scales.

% Second objective: assess if the inclusion of repulsive surface forces changes the character of boundary behaviour?

We have also seen that a change in surface character, from a scenario in the
absence of surface forces to one with a short range surface force, reduces the
prevalence of surface-bound trajectories, with those that would previously
have ended in collision now being reoriented into the deflective regime. This
promotion of deflective behaviour in the presence of additional repulsive
effects is consistent with wild-type promastigotes, where a change in LPG
during metacyclogenesis is thought to result in cell detachment from midgut
epithelium, followed by taxis towards the sandfly foregut \cite{Pimenta1992}.
Indeed, we conjecture that this taxis is aided by the transfer of
promastigotes into the bulk and away from the no-slip boundary, so that cells
are more susceptible to convection by background flows. Such flows may be
associated with sandfly regurgitation, stimulated by the parasite's production
of promastigote secretory gel in their infective form \cite{Bates2007}. Thus
the hydrodynamic interaction behind the deflection of virtual promastigotes
may be responsible for \invivo movement of promastigotes into the bulk, and
could therefore facilitate the transmission of the infective form of the
parasite from vector to host.

However, whilst we have noted that stable boundary swimming of virtual
promastigotes does not occur with a repulsive surface potential, this is not
the case for the human spermatozoon \cite{Ishimoto2014,Smith2009a}. Differing
in both hydrodynamic classification and cell morphology, it is not clear in
the context of the presented results which feature drives the contrasting
behaviours observed when accounting for short range surface forces, as
time-reversibility is lost due to the boundary force. However, the above
results indicate that cell morphology may not have significant impact in
general on stable surface swimming, thus the observed behavioural differences
between virtual promastigotes and human spermatozoa are hypothesised to be
primarily resultant of the contrasting methods of locomotion.

% Extra observation: tip-first alignment to boundary promotes collisions as seen in vivo.

Remarkably, we have observed a morphology-dependent mechanism for the
promotion of tip-first boundary collision for virtual promastigotes near
non-repulsive boundaries. We hypothesise that this mechanism may provide an
explanation for the observed behaviour of \invivo promastigotes, where
flagellum-first attachment is well-documented. This behaviour lacks
an evidenced driving mechanism, though it has been postulated to be due to the
flagellum-first nature of \leish{}swimming \cite{Bates2009}. Here
\insilicons, our study of virtual promastigotes suggests a refined mechanism,
whereby  the comparatively large body size of the promastigote results in the
emergence of drag-based reorientation of notable magnitude, highly dependent
on body lengthscale, aligning the virtual flagellum such that the distal tip
initiates boundary contact.

% Future work
In \Cref{app:morphology} we have examined the behavioural effects of changing
body lengthscale and aspect ratio, demonstrating that reported behaviours are
robust to typical observed variation in these morphological parameters. There
remains significant scope in future work to relax the assumption of body
axisymmetry in order to more accurately model \leish promastigotes, and thus
determine the impact of symmetry-breaking body geometry on boundary
behaviours. One might expect there to be a significant dependence of behaviour
on aspects of cell morphology, with certain realistic \leish promastigote body
curvatures breaking all symmetry and thus adding further complexity to the
dynamics. Additionally, the classification of boundary and general swimming
behaviours for the different morphologies of promastigotes observed in
sandflies, which is more diverse than in culture, is likely to be highly
relevant in the continued study of \leish spp.\ and in the general
investigation into hydrodynamic pullers with flagellum-scale cell bodies.

% Summary

In summary, we have investigated in detail the boundary behaviours of a
flagellated puller, a virtual \lmex promastigote equipped with a determined
planar tip-to-base beat pattern, and have observed that stable boundary
accumulation does not feature amongst the range of exhibited behaviours,
irrespective of the inclusion of repulsive surface forces. Instead, long-time
promastigote behaviour may be divided into two distinct categories based on
initial location and orientation: those that are deflected away from the
boundary, and those that collide tip-first with the boundary. However, whether
or not these behaviours are truly representative of \leish promastigotes in
their microenvironments requires further exploration. Nevertheless, our
results suggest that the observed behaviour in the sandfly vector midgut may
be explained by the hydrodynamic interactions between a promastigote and a
boundary, enabling cell attachment and subsequent detachment at life cycle
stages appropriate for \leish survival and virulence. In particular, we have
seen that boundary collision via the distal tip of the flagellum is promoted
mechanically in virtual promastigotes by a combination of a large cell body
and tip-to-base beating, evidencing a remarkable morphology-dependent
hydrodynamical mechanism of boundary approach.

\section*{Acknowledgements}
B.J.W.\ is supported by the UK Engineering and Physical Sciences Research
Council (EPSRC), grant EP/N509711/1. R.J.W.\ is supported by a Wellcome Trust
Sir Henry Wellcome Fellowship [103261/Z/13/Z] and a Wellcome Trust
Sir Henry Dale Fellowship [211075/Z/18/Z], with equipment supported by a
Wellcome Trust Investigator Award [104627/Z/14/Z]. K.I.\ acknowledges JSPS
Overseas Fellowship and MEXT Leading Initiative for Excellent Young Researcher
(LEADER).

\appendix
% This redefines the thesection command to fix cref for use in appendices.
\gdef\thesection{\Alph{section}} % corrected redefinition of "\thesection"
\makeatletter
\renewcommand\@seccntformat[1]{Appendix \csname the#1\endcsname.\hspace{0.5em}}
\makeatother

%!TEX root=../author_accepted.tex

\section{Incompressible Stokes equations}
\label{app:stokes}
Briefly stated, the dimensional Stokes equations for an incompressible
Newtonian fluid with velocity field $\vec{u}$ and pressure $p$ are given by
\begin{equation}\label{eq:stokes}
	\mu\nabla^2\vec{u} = \nabla p\,, \quad \nabla\cdot\vec{u}=0\,,
\end{equation}
where $\mu$ is the dynamic viscosity of the fluid. We non-dimensionalise with
lengthscale $L=10\si{\um}$ and velocity scale $U=1\si{\um\per\s}$, and scale
pressure with $\mu U/L$, taking the advective timescale $L/U$ and using the
dynamic viscosity of water at 25\si{\celsius}. For typical \lmex values this
gives a Reynolds number on the order of $10^{-3}$, so the inertialess limit of
Stokes equations is appropriate here. Treatment of the non-dimensionalised
equations follows \citet{pozrikidis2002}, using a standard Green's function
approach to form the 3-dimensional boundary integral equation over a
2-dimensional surface, as given in \cref{eq:boundary_integral}. This equation
is then applied to the surface of the virtual promastigote and discretised,
with the resulting linear system being solved for cell velocities
$\vec{U},\vec{\Omega}$ given the prescribed flagellar kinematics, together
with the force and torque conditions on the swimmer.

\section{Meshing the virtual promastigote}
\label{app:mesh}

Both icosahedral and triangulated-cubic meshes were used for the
discretisation of the virtual promastigote body. Subdivision was performed by
the bisection of existing edges, increasing the element count by a factor of 4
per subdivision. For the flagellum, a cylindrical mesh with spherically-capped
ends was used, with the vertices being constructed in equispaced circular
bands and triangular elements constructed between them. Typically in
simulations a mesh of 320 elements was used for the cell body, and
324 for the flagellum. High sensitivity of simulation results on mesh geometry
was noted, with a low resolution symmetry-breaking icosahedral mesh producing
artefacts in long-time simulations. Thus simulation results were verified at
higher mesh resolutions, where artefacts were of negligible magnitude.

\section{Effects of body morphology}
\label{app:morphology}
\setcounter{figure}{0}

\begin{figure*}[t]
	\centering
	\begin{subfigure}[c]{0.4\textwidth}
		\centering
		\includegraphics[width=\textwidth]{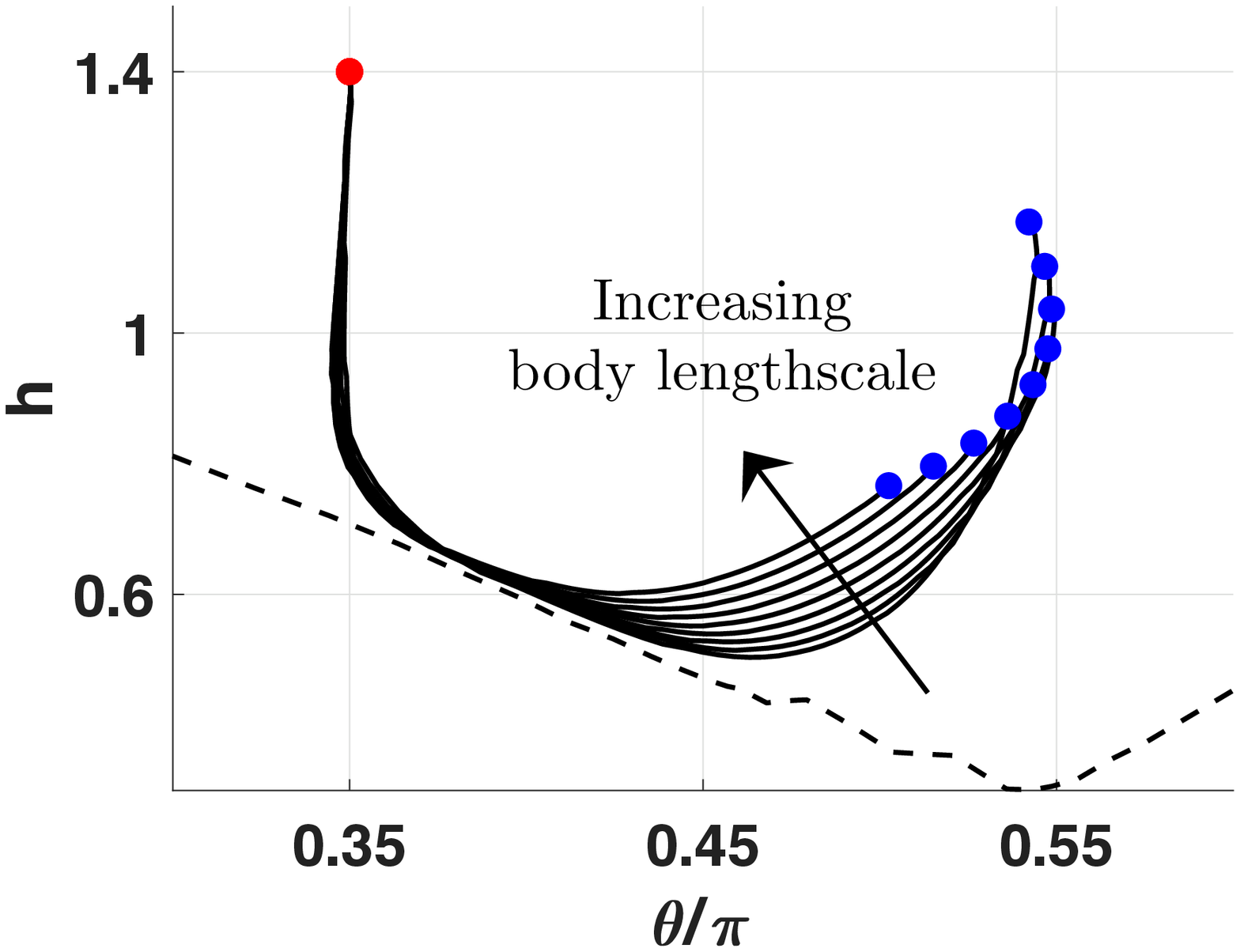}
		\caption{\label{fig:lengthscale_paths_repulsive}}
	\end{subfigure}
	\begin{subfigure}[c]{0.4\textwidth}
		\centering
		\includegraphics[width=\textwidth]{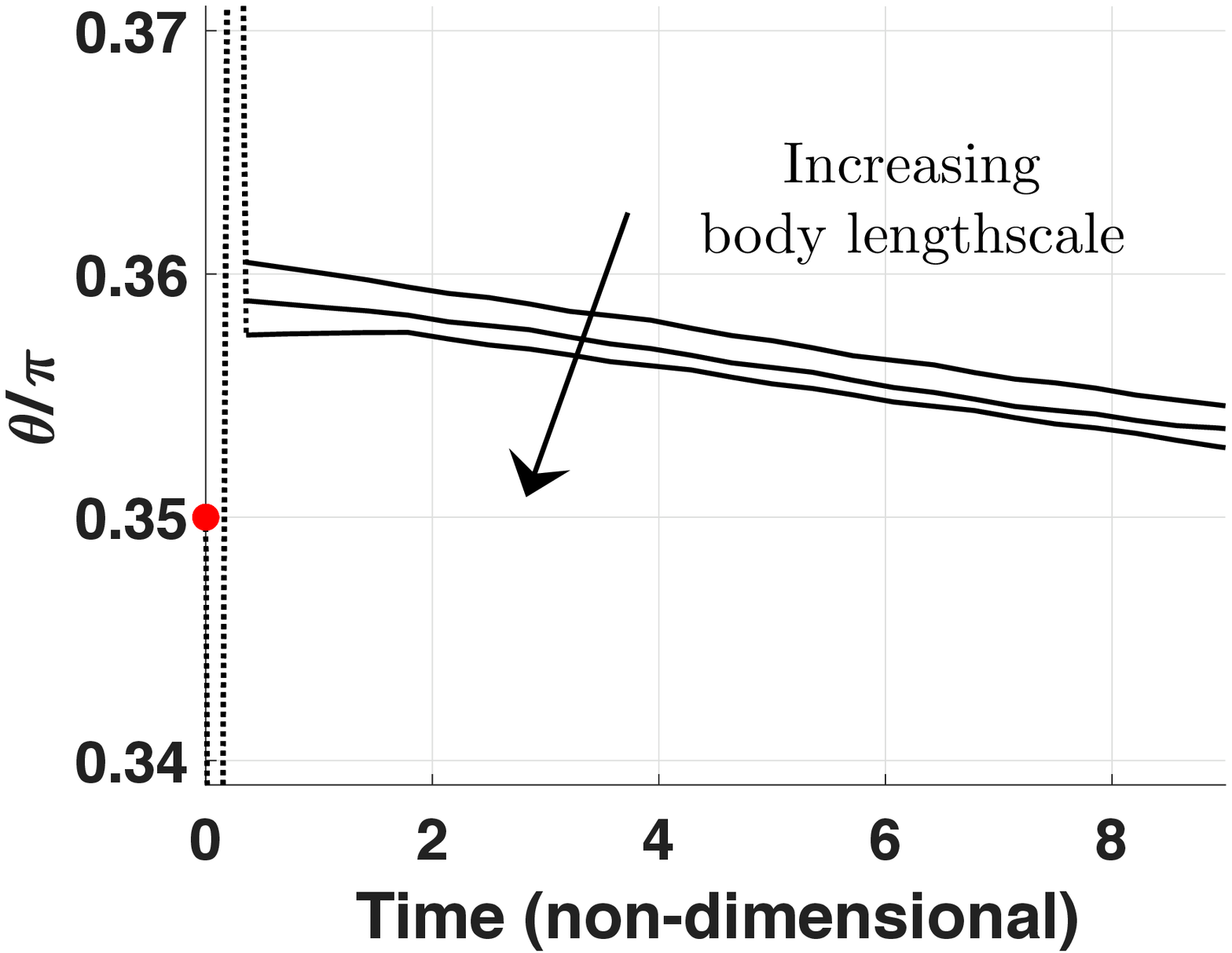}
		\caption{\label{fig:lengthscale_paths_passive}}
	\end{subfigure}
	\caption{Simulated swimming of virtual promastigotes with
	differing body lengthscales. (\subref{fig:lengthscale_paths_repulsive})
	Beginning from a sample initial condition of $(\theta,h)=(0.35\pi,1.4)$,
	with the initial location in phase space being shown in red and the
	simulation endpoints in blue, the smoothed paths of virtual swimmers are
	depicted as black curves. We simulate virtual promastigotes with a
	modified body lengthscale, where the aspect ratio of the body is kept
	fixed and the length of the body is sampled from a range of 7-17\si{\um},
	representing typical variation in
	\leishns. The black dashed line approximately separates off configurations that
	represent intersection with the boundary. We observe
	qualitatively-unchanged swimming behaviour in the presence of a repulsive
	surface potential, with reduced deflection being seen for larger-bodied
	swimmers. (\subref{fig:lengthscale_paths_passive}) Simulating the swimmers
	in the absence of short range boundary forces, halting prior to boundary
	contact, shown is a time series plot of the evolution of $\theta/\pi$,
	with the initial configuration of all cells being
	$(\theta,h)=(0.35\pi,1.4)$ and lengthscales sampled from the same range as
	in (\subref{fig:lengthscale_paths_repulsive}). Greater alignment towards
	the boundary is observed for swimmers with increased body lengthscales.
	The dotted section of the curve shows an initial unsmoothed transient,
	with the common initial value highlighted in red. (Colour online).}
	\label{fig:lengthscale_paths}
\end{figure*}

Given the reported variation in \leish{}promastigotes
\cite{Wheeler2011,Ambit2011}, we examine the effects on swimming of altering
the morphological parameters describing the swimmer body. With the body length
of typical promastigotes varying approximately between 7\si{\um} and
17\si{\um} \cite{Wheeler2011}, we simulate the motion of modified virtual
promastigotes with body lengths sampled from this range. Here we fix the body
aspect ratio to be that of the virtual promastigote, and show sample
kinematics in \cref{fig:lengthscale_paths} for a single initial configuration.
\cref{fig:lengthscale_paths_repulsive} presents the motion of swimmers when a
repulsive boundary force is included. We observe that, despite differing in
body lengthscale, the swimmers follow qualitatively-similar paths in phase
space and exhibit the same overall behaviours. The larger-bodied swimmers are
seen to undergo reduced reorientation and displacement in phase space when
compared with smaller-bodied virtual swimmers, consistent with larger bodies
having increased overall drag as well as increased near-wall hydrodynamic
resistance, the latter as presented in \Cref{tab:drag}. Without repulsive
short range surface forces the same level of qualitative agreement is present,
as evident from the time series of \cref{fig:lengthscale_paths_passive}, with
increased reorientation towards the boundary for larger-bodied swimmers.

Similar consideration of variation in body aspect ratio from that of the
virtual promastigote yields qualitatively-unchanged dynamics, where
additionally we observe that decreases in aspect ratio result in reduced
swimmer velocities and correspondingly shortened trajectories in phase space
for a given simulation time. With the virtual promastigote previously having
had an aspect ratio of $\sim3.14$, here the aspect ratio was sampled from the
range 2-5, capturing the typical variation seen in \leish{}promastigotes
\cite{Wheeler2011}. Hence we have seen that the reported behaviours of virtual
promastigotes are robust to oberved variations in both aspect ratio and body
lengthscale.

\end{document}